\documentclass[apj]{emulateapj}
\shorttitle{RCS2319+00 supercluster}
\shortauthors{Gilbank et al.}

\newcommand\gsim{\gtrsim}
\newcommand\lsim{\lesssim}
\newcommand\kms{km s$^{-1}$}
\newcommand\bgc{B$_{gcR}$}

\begin{document}

\title{A $z=0.9$ supercluster of X-ray luminous, optically-selected, massive galaxy clusters}

\author{David G. Gilbank\altaffilmark{1,2}, H. K. C. Yee\altaffilmark{1},
E. Ellingson\altaffilmark{3}, A. K. Hicks\altaffilmark{4}, M. D. Gladders\altaffilmark{5,6},
L. F. Barrientos\altaffilmark{7}, B. Keeney\altaffilmark{3}
}
 
\altaffiltext{1}{Dept. of Astronomy and Astrophysics, Univ. of Toronto,
  50 St George Street, Toronto, ON, Canada, M5S 3H4; gilbank@astro.utoronto.ca, hyee@astro.utoronto.ca}
\altaffiltext{2}{Astrophysics and Gravitation Group, Dept. of Physics \& Astronomy, Univ. of Waterloo, Waterloo, ON, Canada N2L 3G1; dgilbank@astro.uwaterloo.ca}
\altaffiltext{3}{Center for Astrophysics and Space Astronomy, 389 UCB , Univ. of Colorado, Boulder, CO  80309; erica.ellingson@colorado.edu, keeney@casa.colorado.edu}
 \altaffiltext{4}{Dept. of Astronomy, Univ. of Virginia, P.O. Box 400325, Charlottesville, VA 
22904; ahicks@alum.mit.edu} 
\altaffiltext{5}{Dept. of Astronomy and Astrophysics, Univ. of Chicago, 5640 S. Ellis Ave., Chicago, IL, 60637; gladders@oddjob.uchicago.edu}
\altaffiltext{6}{Visiting Associate, The Observatories of the Carnegie Institution of
Washington, 813 Santa Barbara St., Pasadena, CA 91101}
\altaffiltext{7}{Departamento de Astronom\'{\i}a y Astrof\'{\i}sica,
Universidad Cat\'{o}lica de Chile, Casilla 306, Santiago 22, Chile; barrientos@astro.puc.cl}

\begin{abstract}
We report the discovery of a compact supercluster structure at z$=$0.9.  The structure comprises three optically-selected clusters, all of which are detected in X-rays and spectroscopically confirmed to lie at the same redshift.  The {\it Chandra} X-ray temperatures imply individual masses of $\sim 5 \times 10^{14} M_\odot$.  The X-ray masses are consistent with those inferred from optical--X-ray scaling relations established at lower redshift.  A strongly-lensed z$\sim$4 Lyman break galaxy behind one of the clusters allows a strong-lensing mass to be estimated for this cluster, which is in good agreement with the X-ray measurement.  Optical spectroscopy of this cluster gives a dynamical mass in good agreement with the other independent mass estimates.  The three components of the RCS2319+00 supercluster are separated from their nearest neighbor by a mere $<3$ Mpc in the plane of the sky and likely $<10$ Mpc along the line-of-sight, and we interpret this structure as the high-redshift antecedent of massive ($\sim10^{15}$ M$_\odot$) z$\sim$0.5 clusters such as MS0451.5-0305.  
\end{abstract}


\section{Introduction}

The cluster RCS231953+0038.0 was discovered in the first Red-Sequence Cluster Survey (RCS-1, \citealt{Gladders:2005oi}) and displays spectacular strong lensing features \citep{2003ApJ...593...48G}.  As such, it has been the subject of extensive follow-up, including multi-object spectroscopy (MOS) \citep{felipe07} and {\it Chandra} X-ray imaging \citep{hicks07}. The {\it Chandra} X-ray observations were targeted so as to include two other cluster candidates (RCS231848+0030.1 and RCS232002+0033.4) which the RCS catalog indicated were consistent with being at the same photometric redshift as the primary cluster target.  All three systems were found to be coincident with extended X-ray emission \citep{hicks07}.  In this Letter we present new follow-up spectroscopy of these three clusters, showing that all three are located at the same redshift and therefore likely physically associated.  Hereafter we refer to these collectively as the supercluster RCS2319+00.  

Throughout, we assume a concordance cosmology with $\Omega_m=0.3$, $\Omega_\Lambda=0.7$, $H_0=70$ \kms~Mpc$^{-1}$.

\section{The optical spectroscopic data}

\begin{figure}
\plotone{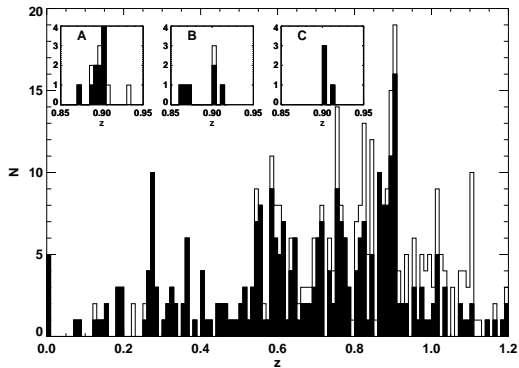}
\caption{Redshift histograms for the full field (27\arcmin~diameter, 12.6 Mpc at z$=$0.9).  The open histogram shows redshifts of all quality and the filled histogram shows only those of higher quality.  A bin size of 0.01 is used which corresponds to $\sim$1500 \kms~ at z$=$0.9.  Insets show redshifts within 1.5 Mpc of each cluster center.}
\label{fig:fullhist}
\end{figure}

Wide-field MOS spanning a field of 27\arcmin~diameter using IMACS on the 6.5-m Baade telescope was obtained on the night of 2006 October 28-29.  The observations consisted of 4 $\times$ 1800s exposures of a single mask comprising 725 slits utilizing a custom band-limiting filter, which restricted the wavelength coverage to 6050-8580\AA, allowing coverage of the key spectral features from [O{\sc ii}]$\lambda$3727 to G-band ($\sim4304$\AA) and somewhat beyond,  at the redshift of the cluster.   The spectra were reduced following a technique similar to that described in \citet{gilbank:07a}.  Briefly, we used the COSMOS software package\footnote{See http://www.ociw.edu/$\sim$oemler/COSMOS.html} to process the 2D spectra.  Our procedure required the additional step of masking second and third order bright skyline contamination, which overlapped many of our first order science spectra due to the highly multi-tiered nature of our data.  The spectra were extracted to 1D using the apall task in IRAF.  Redshifts were measured  using the IRAF task {\sc rvsao}.  All the spectra were visually inspected and assigned redshift confidences based on the system described in \citet{gilbank:07a}.  Hereafter we group redshifts with quality flags 1-3 into a higher confidence category (302 objects) and denote class 4 redshifts as lower confidence (113 objects).  The redshift histograms for all galaxies in this field are shown in Fig.~\ref{fig:fullhist}.  Also included are 16 redshifts centered on RCS231953+0038.0 from VLT spectroscopy \citep{felipe07}.  A clear peak is seen at z$\sim$0.9.  The insets show histograms within 3.2 arcmins (1.5 Mpc) of each cluster center.

\begin{figure*}
\plotone{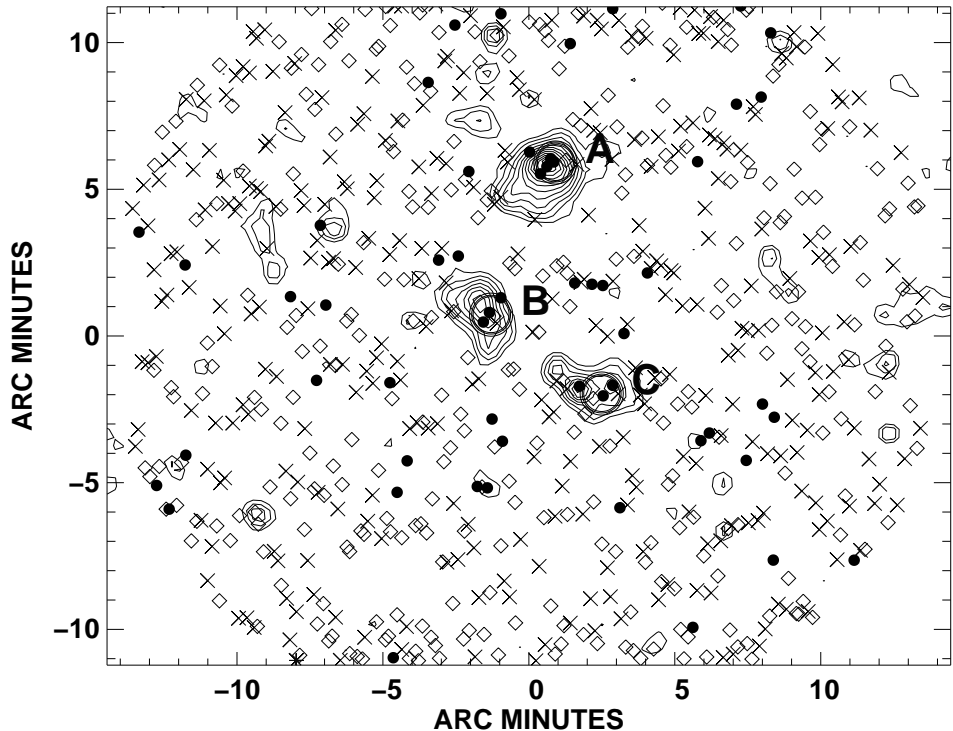}
\caption{Distribution of galaxies in the supercluster structure within the IMACS field. Filled circles show galaxies with IMACS spectroscopic redshifts (both higher and lower confidence) consistent with belonging to the z$=$0.9 structure; crosses are objects inconsistent with this redshift and diamonds denote slits which failed to yield a redshift.  Contours show the significance of red-sequence structures, with levels running from 2.0$\sigma$ in 0.3$\sigma$ intervals.  Note the striking tendency for the major axis of each cluster, as traced by red galaxies,  to point toward its nearest neighbor.  Large open circles indicate X-ray centroids from \citet{hicks07}. 
\label{fig:sigmap}}
\end{figure*}

The properties of the three clusters are summarized in Table \ref{table:props} and we refer to the three clusters as A, B, C, as denoted in the table. Fig.~\ref{fig:sigmap} shows the sky distribution of all spectroscopically observed galaxies within the IMACS field.   Although the field is sparsely sampled, overdensities of spectroscopic members can be seen around the locations of the most significant peaks in the RCS significance map (see \citealt{Gladders:2005oi}).  The visually-identified BCG in each case has a confident redshift near z$=$0.90 (0.9008, 0.9025 and 0.9005 respectively).  

\begin{deluxetable*}{lccccc}
\tablecaption{Summary of clusters\label{table:props}}
\tablewidth{0pt}
\tablehead{
\colhead{Cluster} & \colhead{z$_{spec}$} & \colhead{\bgc~$(h_{50}^{-1} $Mpc$)^{1.77}$} & \colhead{L$_X (10^{44}$ erg s$^{-1}$)} & \colhead{T$_X$ (keV)} & \colhead{M$_{X tot}$ ($10^{14}$ M$_\odot$)} \\
}
\startdata
(A) RCS231953+0038.0 & 0.8972 & $1515\pm280$  & 7.6$^{+0.6}_{-0.4}$ & 6.2$^{+0.9}_{-0.8}$ & 6.4$^{+1.0}_{-0.9}$\\
(B) RCS231948+0030.1 & 0.9024 & $1150\pm320$  & 3.6$^{+0.6}_{-0.4}$ & 6.5$^{+1}_{-1}$  & 5.1$^{+0.8}_{-0.8}$\\
(C) RCS232002+0033.4 & 0.9045 &  $580\pm200$   & 4.2$^{+0.5}_{-0.3}$ & 5.9$^{+2}_{-1}$ & 4.7$^{+0.9}_{-1.4}$\\
\enddata
\tablecomments{An abbreviated name (A-C) is given in addition to each cluster's entry in the RCS catalog; $z_{spec}$ denotes clipped mean redshift within 1.5 Mpc of cluster center; \bgc~is optical richness \citep{Gladders:2005oi}; bolometric luminosities and temperatures are determined within $r_{2500}$, and masses are extrapolated to $r_{200}$}
\end{deluxetable*}

\subsection{Spectroscopy of the gravitational arc}
\begin{figure}
\plotone{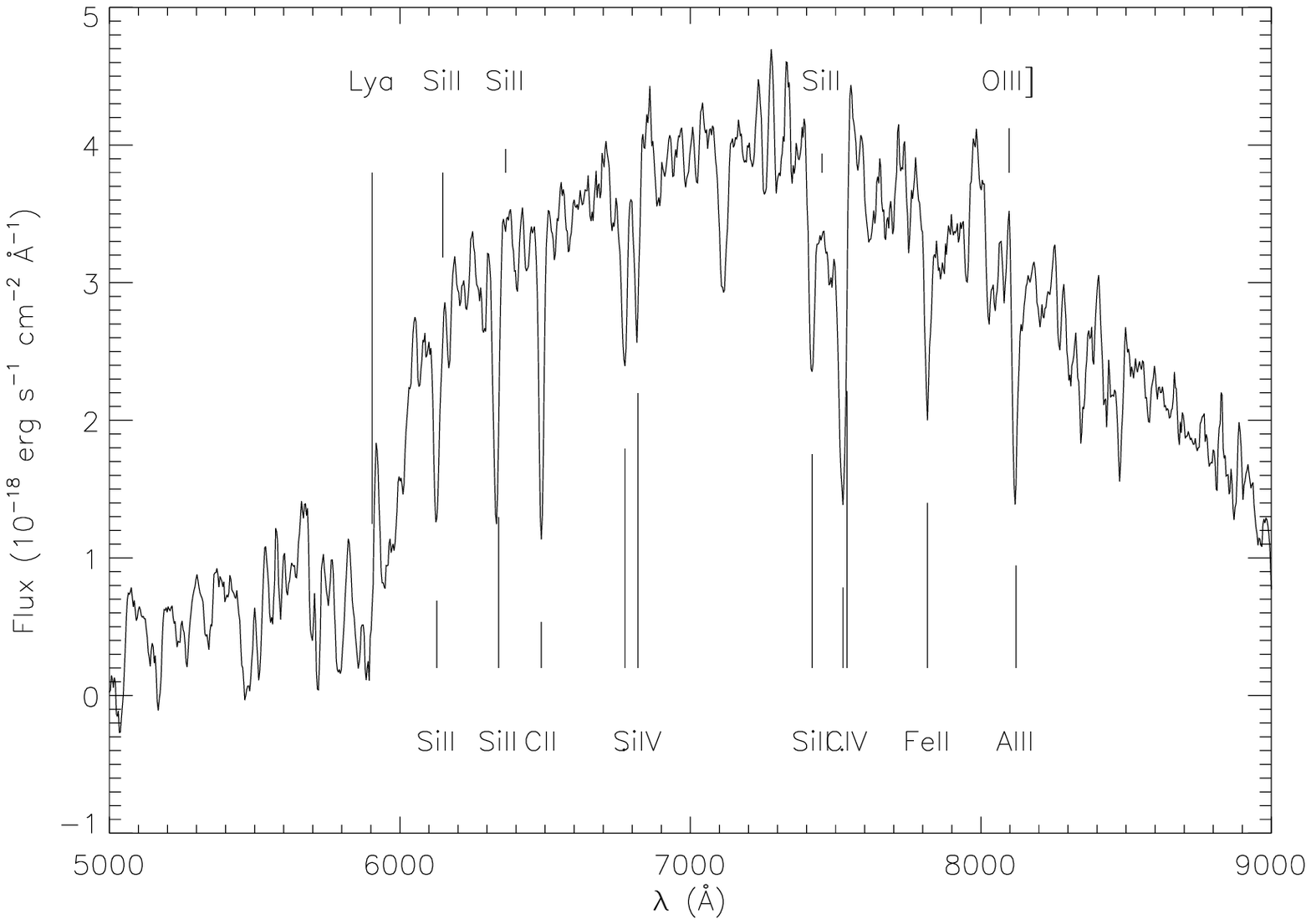}
\caption{GMOS nod-and-shuffle spectrum of the strongly-lensed z$=$3.8605 Lyman break galaxy behind RCS231953+0038. The strongest rest-frame UV absorption and emission lines are overplotted.  Note that Ly$\alpha$ is offset $\sim$900 \kms with respect to the systemic redshift.
\label{fig:arc}}
\end{figure}

Cluster A exhibits several strong lensing features \citep{2003ApJ...593...48G}.  We obtained deep spectroscopy of the $B$-band dropout arc (fig.~2c, \citealt{2003ApJ...593...48G}) in 2003 September-October using GMOS-N on Gemini-North (program ID: GN-2003-Q-19). The target was observed in nod-and-shuffle mode for a total integration time of 9.6 hours and the data reduced as for the Gemini data described in \citet{gilbank:07a}.  The spectrum of the arc is shown in Fig.~\ref{fig:arc} and is that of a high-redshift Lyman-break galaxy (LBG).   As discussed in \citet{Shapley:2003yq}, LBGs frequently display strong outflows, and so the measurement of a redshift is complicated by different spectral features tracing kinematically distinct components of the galaxy.  In order to determine a systemic redshift, we use only stellar photospheric lines and plot their positions at different trial redshifts over the spectrum.  We find a redshift of z$=$3.8605.   Other common lines from interstellar absorption and nebular emission, etc.\ are shown in Fig.~\ref{fig:arc}.  Ly$\alpha$ is seen offset to higher redshift ($\sim$900 \kms) with respect to the systemic velocity, as is often observed \citep{Shapley:2003yq}.

\section{Mass estimates}
\subsection{X-ray masses}

The X-ray observations and analysis are described in \citet{hicks07} and are derived from four {\it Chandra} observations taken over the period 2005 September 28 - October 23, resulting in a total exposure of 74,539 seconds.  Briefly, X-ray temperatures were measured within $r_{\rm 2500}$ and used to derive total masses by extrapolating to $r_{200}$.  The X-ray masses for all three clusters are given in Table~\ref{table:props} and each is around 5$\times$10$^{14}M_\odot$. 

\subsection{Dynamical mass for cluster A}
Cluster A has sufficient spectroscopic members (by including the additional VLT data) to measure an approximate velocity dispersion.  12 redshift class 1-4 members yield $\sigma=(990\pm240)$ \kms~or 9 higher confidence (class 1-3) members yield $\sigma=(790\pm200)$ \kms.  These velocity dispersions would correspond to virial mass estimates \citep{carlberg97} of $6.1^{+5.6}_{-3.4} \times 10^{14}$ M$_\odot$ and $3.1^{+3.0}_{-1.8} \times 10^{14}$ M$_\odot$, which are in good agreement with the X-ray derived mass of $6.4^{+0.6}_{-0.6} \times 10^{14}$ M$_\odot$, within the broad uncertainties of the dynamical estimate.  Using the richness -- dynamical mass relation established at lower-redshift \citep{blindert07a}, the \bgc~values of clusters A-C would correspond to expected masses of M$=(9.5, 5.4, 1.3) \times$ 10$^{14}$ M$_\odot$ respectively.  The 1$\sigma$ intrinsic scatter in the observed relation is $\sim$0.7 dex \citep{blindert07a} and thus the inferred masses are consistent with the measured X-ray masses within this broad scatter.

\subsection{Strong lensing mass for cluster A}
We can use the radial distance and redshift of the strongly-lensed z$=$3.8605 galaxy to make an estimate of the mass of cluster A within this radius.  A simple circular fit to the position of the arc with respect to the brightest cluster galaxy (BCG) shows that it lies at a radius of $\sim$12\arcsec~ (0.1 Mpc).  Setting this equal to the Einstein radius and assuming that the density profile of the cluster is that of an isothermal sphere would imply a central velocity dispersion of 860 \kms [using $\theta_E = (4\pi^2\sigma^2/c^2)(D_{ds }/D_{os})$, e.g., \citet{sef:1992kk}].  Using $M = 7.3 \times 10^{12} h_{100}^{-1} M_\odot (\sigma /100$ km s$^{-1})^2 (r/1$ Mpc$)$ \citep{Hoekstra:2003ky} gives an estimate of the mass enclosed within this radius as $4.8 \times 10^{13} M_\odot$.  Recalculating the X-ray mass within a cylinder of radius 0.1 Mpc, to be directly comparable to the lensing mass estimate, gives a mass of $M_X = (4.4\pm0.4) \times 10^{13} M_\odot$, in excellent agreement with that inferred from the simple strong lensing estimate.

\section{The cosmological significance of this system}

The separations of the components of this system, both in the plane of the sky ($<$ 3 Mpc) and along the line of sight ($\sim$10 Mpc, assuming that the velocity difference is due to Hubble flow) are significantly closer than those seen in other supercluster candidates at these redshifts \citep{Lubin:2000or,Swinbank:2007eg}.  This invites the question: will these merge to form a more massive cluster?  To answer this, we consider a simple estimate \citep{Sarazin:2002rv}.  If we assume that the clusters are not moving apart with the Hubble flow and that their relative velocity is comparable to their restframe LOS velocity difference ($\sim$1200 \kms), then the time taken to traverse the $\sim$3 Mpc separation would imply that the individual cluster components would merge by z$\sim$0.5.  The merger redshift is also comparable if we assume that the infall velocity is of the order of the velocity dispersion of cluster A.  Thus, this system would evolve into a cluster of mass $\gsim$ 10$^{15}$ M$_\odot$ 
at z$\sim$0.5, similar to the z$=$0.54 cluster MS0451.5-0305 \citep{Donahue:1996uo,Ellingson:1998tt}.  

We explore the expected space density of such systems by utilizing the cluster catalogs from the light cone output of the Virgo Consortium Hubble Volume simulation\footnote{See http://www.mpa-garching.mpg.de/NumCos} \citep{Evrard:2002mx}.  Since the abundance of superclusters depends sensitively on the mass threshold chosen, we adopt the conservative approach of searching for clusters of 3.3$\times$10$^{14}$ M$_\odot$ (the 1$\sigma$ lower limit for the cluster with the lowest X-ray mass) within similar separations.  In one octant of the sky, we find only one such system in the redshift range 0.8 to 1.0, corresponding to a surface density of $1.9 \times 10^{-4}$ per sq. deg.  We should therefore have a probability of $<$2\% of finding such an object in the 92 deg$^2$ of RCS-1, if the Hubble Volume simulation gives an accurate description of the abundance of structure.  Relaxing the maximum separation requirement slightly to 10 Mpc (which is the maximum inferred spatial separation if the LOS velocity difference is due to cosmic expansion), we find five such triple clusters, corresponding to a surface density of $9.5 \times 10^{-4}$ per sq. deg. or a probability of $<$9\% of finding such an object in the RCS-1 survey area.  If the clusters are in fact less massive than our mass estimates suggest, then the probability of finding such a supercluster structure increases.  We repeat our analysis using different mass thresholds for the component clusters until the surface density of such structures approaches that of one system per RCS-1 survey area.  To do this requires the individual clusters to have masses of only $\sim 2.3 \times 10^{14}$ M$_\odot$, i.e., {\it all three} would have to be at least a factor of two less massive than our best estimates, which corresponds to a $>$2$\sigma$ deviation.  The Hubble Volume simulation was performed with a value for the amplitude of fluctuations, $\sigma_8$, of 0.90, whereas recent measurements such as the WMAP 3-year results \citep{Spergel:2007fw}, find a lower value ($\sigma_8=0.761^{+0.049}_{-0.048}$).  Lowering $\sigma_8$ lowers the abundance of massive structures, particularly at higher redshift, making the chances of finding such a supercluster more unlikely in this cosmology.

\section{Discussion and Conclusions}

We have presented spectroscopic confirmation of three clusters, selected from a large optical survey and found to be X-ray luminous from {\it Chandra} observations.  Two independent mass estimators for one cluster (including a strong-lensing estimate from a newly confirmed z$\sim$4 Lyman break galaxy behind the cluster) support the X-ray mass estimates. 

The major axes of the three components of supercluster RCS2319+00, as traced by the red-sequence significance \citep{Gladders:2005oi} contours, all point towards the nearest-neighbor cluster (Fig.~\ref{fig:sigmap}).  This alignment is also largely mirrored by the X-ray contours (fig.~1, \citealt{hicks07}).  The tendency for neighboring clusters to point toward each other was first noticed by \citet{Binggeli:1982ok} and is further independent evidence that these clusters are associated.  In addition, this may give some clues as to the 3D alignment of the clusters' haloes.  One of the interpretations for the high incidence of strong-lensing clusters in RCS-1 is that the strongly-lensing systems may have their major axes directed along the line-of-sight \citep{2003ApJ...593...48G}.  The alignment with the neighboring cluster, coupled with the small line-of-sight separation, might suggest that this is not the explanation in this case.  Detailed lens modeling will be required to try to constrain halo orientation and is beyond the scope of this Letter. 

The supercluster RCS2319+00 comprises three clusters, each massive in their own right ($\sim 5 \times 10^{14}$ M$_\odot$), which are likely to merge by low-redshift to form a system $> 1.5 \times 10^{15}$ M$_\odot$.  Thus, this system gives us a unique opportunity to study one of the most massive clusters in the universe in the process of assembly.  Two other groups \citep{Lubin:2000or,Swinbank:2007eg} have reported finding superclusters at similar redshifts, but these represent somewhat different systems to that presented here.  The \citet{Lubin:2000or} supercluster (Cl1604) comprises at least four components \citep{Gal:2005wf} spanning $\sim10$ Mpc by $\sim100$ Mpc ($0.865 < z< 0.921$).  The clusters range in velocity dispersion from $<500$ \kms to $\sim$900 \kms. However, the X-ray temperature for the main cluster of Cl1604 is only 2.5 keV \citep{Lubin:2004no}, suggesting that at least this component is less massive than its velocity dispersion would imply, possibly due to the cluster being dynamically unrelaxed. The lower masses and larger separations suggest that Cl1604 is less likely to merge into as massive a cluster as RCS2319+00.  The \citet{Swinbank:2007eg} supercluster comprises five systems with approximate velocity dispersions $\lsim500$ \kms~($M \lsim 8 \times 10^{13}$ M$_\odot$) spread across 30 Mpc.  The lower masses of these systems is not surprising given the low space density of massive clusters and the 1 sq. deg. size of their survey.  
 
The chances of finding a structure as massive as RCS2319+00 within the RCS-1 survey are $<9\%$ with $\sigma_8=0.90$ and less with a lower value, using the currently favored cosmological parameters.  This may indicate that refinements to the current cosmological model are still to be made, but we are, of course, cautious about using the existence of a single system to make statements about cosmology.  The order-of-magnitude greater area of RCS-2 \citep{Yee:2007qb} will allow us to place stronger constraints on the abundance of such massive structures.  

A program of detailed wide-field spectroscopy to fully explore the 2319+00 structure is underway.  In conjunction with our comprehensive on-going IMACS spectroscopy of a core sample of RCS clusters, selected to uniformly sample mass and redshift space, we will be able to probe the processes which drive galaxy evolution in clusters as a function of cluster mass.  As an unusually massive cluster in the early stages of merging, and through comparison with lower-redshift $\sim10^{15}$ M$_\odot$ clusters, RCS2319+00 may give us a unique opportunity to identify processes which occur in massive clusters prior to their final assembly.

\acknowledgements
This paper includes data gathered with the 6.5 meter Magellan Telescopes located at Las Campanas Observatory, Chile.  Based on observations obtained at the Gemini Observatory, which is operated by the Association of Universities for Research in Astronomy, Inc., under a cooperative agreement with the NSF on behalf of the Gemini partnership: the National Science Foundation (United States), the Science and Technology Facilities Council (United Kingdom), the National Research Council (Canada), CONICYT (Chile), the Australian Research Council (Australia), CNPq (Brazil) and SECYT (Argentina).  The RCS project is supported by grants to H.K.C.Y. from the Canada Research Chair Program,  the Natural Sciences and Engineering Research Council of Canada (NSERC) and the University of Toronto.  E.E. acknowledges NSF grant  AST-0206154.  M.D.G. acknowledges partial support for this work provided by NASA through Hubble Fellowship grant HF-01184.01 awarded by the Space Telescope Science Institute, which is operated by the Association of Universities for Research in Astronomy, Inc., for NASA, under contract NAS 5-26555.  L.F.B. acknowledges the support of the FONDAP center for Astrophysics and CONICYT under proyecto  FONDECYT 1040423.

\end{document}